\documentclass[
prl,
notitlepage,
superscriptaddress,
preprintnumbers,
twocolumn,
footinbib,
amsmath,amssymb,
aps,
]{revtex4-2}

\usepackage{graphicx}
\usepackage{dcolumn}
\usepackage{bm}
\usepackage{hyperref}

\usepackage{comment}
\usepackage{subcaption}
\usepackage{booktabs}
\usepackage{color}
\usepackage{mathtools}
\usepackage[normalem]{ulem} 
\usepackage{amsmath} 
\usepackage{orcidlink}
\usepackage{makecell}

\usepackage{url}
\hypersetup{colorlinks=true, citecolor=cyan}

\newcommand{\be}{\begin{equation}}
\newcommand{\ee}{\end{equation}}

\newcommand{\secl}[1]{\noindent \textit{\textbf{#1—}}}

\definecolor{mygreen}{RGB}{0,130,0}

\newcommand*{\diff}{\,\mathrm{d}}

\begin{document}

\preprint{RESCEU-26/26}

\author{Daiki Watarai\orcidlink{0009-0002-7569-5823}}
\email{dwatarai@caltech.edu}
\affiliation{Research Center for the Early Universe (RESCEU), Graduate School of Science, The University of Tokyo, Tokyo 113-0033, Japan}
\affiliation{TAPIR, California Institute of Technology, Pasadena, CA 91125, USA}

\author{Hayato Imafuku\orcidlink{0009-0001-3490-8063}}
\email{imafuku-hayato@g.ecc.u-tokyo.ac.jp}
\affiliation{Graduate School of Science, The University of Tokyo, Tokyo 113-0033, Japan}
\affiliation{Research Center for the Early Universe (RESCEU), Graduate School of Science, The University of Tokyo, Tokyo 113-0033, Japan}

\author{Fumihiro Chuman\orcidlink{0009-0004-7714-5395}}
\email{fschumann@chiba-u.jp}
\affiliation{Department of Physics, Graduate School of Science, Chiba University, 1-33 Yayoicho, Inage, Chiba 263-8522, Japan}

\date{\today}

\begin{abstract}
Galaxy clustering encodes information about the host galaxies of binary black hole (BBH) mergers even when individual hosts remain unidentified, because the distribution of neighboring galaxies depends statistically on the host-galaxy population.
We construct event-level host-type likelihoods from host-conditioned galaxy-count distributions within gravitational-wave localization regions and combine them to infer the fractions of BBH mergers hosted by different galaxy populations.
The framework applies to common and rare galaxy populations without prescribing formation efficiencies or merger delay-time distributions.
Using measured star-forming and quenched galaxy clustering, idealized one-year forecasts assuming all mergers occur in quenched galaxies yield median one-sided 90\% lower credible limits of 0.19, 0.48, and 0.92 on the quenched-host fraction for O5, post-O5, and third-generation networks.
These forecasts illustrate the potential for host-galaxy population inference before the third-generation detector era.
\end{abstract}

\title{Galaxy Clustering Can Reveal Host-Galaxy Properties of Binary Black Hole Mergers}
\maketitle

\secl{Introduction}
\label{sec:intro} 
The growing sample of gravitational-wave (GW) detections by the LIGO-Virgo-KAGRA collaboration provides an opportunity to investigate the astrophysical origin of binary black hole (BBH) mergers (e.g.,~\cite{LIGOScientific:2026wfs,2026arXiv260527226T}). 
Their host galaxies and surrounding large-scale environments provide complementary information about BBH formation histories beyond BH masses, spins, and merger rates inferred from GW parameter estimation. 
However, BBH mergers generally lack identified electromagnetic counterparts, and their GW localization volumes typically contain many candidate host galaxies. 
What can we learn about BBH merger host galaxies and their environments directly from observations, without identifying individual hosts or assuming models of BBH formation and host-dependent merger rates?

Galaxy clustering provides a way to address this question, since the distribution of neighboring galaxies around a host galaxy depends statistically on the properties of the host.
These host-dependent differences are expected to be less diluted by averaging as GW localization scales approach the characteristic scales of galaxy clustering.
Galaxy counts within a localization volume may therefore carry statistical information not only about the large-scale environment traced by a BBH merger but also about the properties of its host galaxy, even when multiple candidate hosts remain.

Where BBHs merge will help reveal how they formed. 
Models linking BBH formation to galaxy evolution predict host-galaxy populations shaped by metallicity-dependent formation, merger delays, and galaxy assembly~\cite{Mapelli:2017hqk,Cao:2017ztl,Artale:2019doq,Santoliquido:2022kyu,Srinivasan:2023vaa,Rauf:2023oiy}. 
These host-galaxy preferences also affect the predicted clustering of BBH mergers and their large-scale environments~\cite{Scelfo:2020jyw,Peron:2023zae,Smith:2025bkq,Hosseini:2026eeq}. 
Measuring host-galaxy properties and environments without imposing these formation and host-dependent merger-rate prescriptions therefore provides an independent test of formation scenarios, complementing constraints from BH masses, spins, and merger-rate evolution.

Several approaches combine GW observations with information about galaxy populations to constrain BBH merger host galaxies.
GW--galaxy cross-correlations probe the large-scale environments traced by mergers (e.g.,~\cite{Raccanelli:2016cud,Scelfo:2018sny,Calore:2020bpd,Zazzera:2024agl,Chuman:2025ezp}) and can be used to connect GW clustering to host-galaxy properties and formation scenarios~\cite{Hosseini:2025plz,Moncrieff:2025xle}.
The observed redshift evolution of the BBH merger rate provides complementary information on likely host-galaxy properties~\cite{Vijaykumar:2023bgs}, as well as progenitor formation histories and merger delay-time distributions~\cite{Schiebelbein-Zwack:2026sqj}.
Spatial associations between GW localization volumes and catalogs of rare tracers, such as active galactic nuclei (AGN), have also been used to constrain the fraction of BBH mergers hosted by these populations without identifying individual hosts (e.g.,~\cite{Bartos:2017ggb,Veronesi:2022hql,Veronesi:2023ugk,Veronesi:2024tef,Zhu:2025rdv,Pouw:2026lar}).
In GW cosmology, dark-siren analyses incorporate candidate-host information and galaxy properties~\cite{LIGOScientific:2018gmd,Hanselman:2024hqy,Li:2025hrh,Alfradique:2025tbj}, with recent studies jointly inferring cosmology and host-galaxy
weighting~\cite{Gray:2026loe,Agapito:2026pbb} and further obtaining posterior probabilities for individual candidate hosts~\cite{Agapito:2026pbb}.

In this \textit{Letter}, we develop an observation-driven framework that uses the galaxy environment of each GW event to infer its host-galaxy type, even when multiple candidate hosts remain.
We describe the abundance and composition of neighboring galaxies through host-conditioned count distributions, which can be calibrated from observed or mock catalogs for both common galaxy populations and rare tracers.
The resulting event-level likelihoods quantify the relative support for different host-galaxy types and combine to constrain the fractions of BBH mergers hosted by different galaxy populations.
Informative individual events can therefore constrain host-galaxy types without requiring an ensemble measurement of GW--galaxy cross-correlations.
Given these distributions, the inference requires no prescribed BBH formation efficiencies, merger delay-time distributions, or mapping from galaxy properties to BBH merger rates.

We apply the framework to star-forming (SF) and quenched (Q) galaxies, whose relative contributions as the host galaxies probe the connection between mergers and host star-formation activity. 
Using their measured clustering, we quantify both the localization scales relevant to individual host-galaxy identification and the information about host-galaxy types available when multiple candidate galaxies remain. 
Combining this analysis with localization forecasts for O5, post-O5, and third-generation GW detector networks, we assess how information from individual events accumulates to constrain the fraction of BBH mergers hosted by Q galaxies.
Throughout this work, we assume a flat $\Lambda$ cold dark matter cosmology with $\Omega_{\rm m}=0.27$, $\Omega_\Lambda=0.73$, and $H_0=70\,{\rm km\,s^{-1}\,Mpc^{-1}}$.

\secl{Host-conditioned galaxy counts} 
\label{sec:galaxy_clustering} 
Galaxy populations with different clustering properties have different distributions of neighboring galaxies, providing statistical information about the host-galaxy type of a BBH merger even when the individual host remains unidentified.

Let $a$ and $b$ denote two disjoint galaxy populations that together exhaust the galaxy sample, with mean number densities $\bar n_a$ and $\bar n_b$.
We consider $(a,b)=({\rm Q},{\rm SF})$ below, but the formulation also applies to other binary classifications based on stellar mass, metallicity, morphology, AGN activity, or galaxy-cluster membership.
Assuming statistical homogeneity and isotropy, let $\xi_{h,h'}(r)$ denote the real-space two-point auto- or cross-correlation function between distinct galaxies of types $h,h'\in\{a,b\}$ separated by comoving distance $r$.

We characterize galaxy environments through counts in cylinders~\cite{Reid:2008sy,2011ApJ...726....1B,Oguri:2014dqa} centered on a conditioning galaxy.
Each cylinder has transverse comoving radius $R_\perp$ and line-of-sight comoving half-length $L_\parallel$.
Its geometry is specified by $\boldsymbol{\theta}^{\rm cyl}:=(R_\perp,L_\parallel)$, with aspect ratio $A:=L_\parallel/R_\perp$ and volume $V(\boldsymbol{\theta}^{\rm cyl})=2\pi R_\perp^2L_\parallel$.
Following the counts-of-neighbors formalism \citep{1980lssu.book.....P}, the mean number of neighboring galaxies of type $h'$ around a galaxy of type $h$ is
\begin{align}
\label{eq:N_bar}
    \overline{N}_{h'|h}(\boldsymbol{\theta}^{\rm cyl})
    &=\bar n_{h'}\int_V \diff^3\boldsymbol r\,
    [1+\xi_{h,h'}(r)]
    \nonumber\\
    &=\overline{N}_{h'}(\boldsymbol{\theta}^{\rm cyl})
    [1+\overline{\xi}_{h,h'}(\boldsymbol{\theta}^{\rm cyl})]\:,
\end{align}
where $\overline{N}_{h'}=\bar n_{h'}V$ is the unconditional mean count and $\overline{\xi}_{h,h'}:=V^{-1}\int_V\diff^3\boldsymbol r\, \xi_{h,h'}(r)$ is the correlation function averaged over separations from the central galaxy within the cylinder.

For BBH mergers, we identify the conditioning galaxy with the host and adopt the host-centered cylinder as an idealized description of a GW localization volume.
We assume that the host belongs to the galaxy sample and that, within each population, BBH merger hosts share the clustering properties of that population as a whole.
Under the hypothesis $\mathcal H_h$ that the host belongs to population $h$, including the host itself gives the mean total count $\mu_{h'}^{(h)}=\delta_{h,h'}+\overline N_{h'|h}$, where $\delta_{h,h'}$ is the Kronecker delta. 
For the count vector
\begin{equation}
\label{eq:count_vector}
    \boldsymbol{N}(\boldsymbol{\theta}^{\rm cyl})
    :=\begin{pmatrix}
        N_a(\boldsymbol{\theta}^{\rm cyl})\\
        N_b(\boldsymbol{\theta}^{\rm cyl})
    \end{pmatrix}\:,
\end{equation}
the conditional means
$\boldsymbol{\mu}^{(h)}(\boldsymbol{\theta}^{\rm cyl})
:=\langle\boldsymbol{N}(\boldsymbol{\theta}^{\rm cyl})
\mid\mathcal{H}_h\rangle$ are therefore
\begin{align}
\label{eq:mean_count_a}
    \boldsymbol{\mu}^{(a)}(\boldsymbol{\theta}^{\rm cyl}) =
        \begin{pmatrix}
        1+\overline{N}_{a|a}(\boldsymbol{\theta}^{\rm cyl})\\
        \overline{N}_{b|a}(\boldsymbol{\theta}^{\rm cyl})
        \end{pmatrix}\:,\\
\label{eq:mean_count_b}
    \boldsymbol{\mu}^{(b)}(\boldsymbol{\theta}^{\rm cyl}) =
        \begin{pmatrix}
        \overline{N}_{a|b}(\boldsymbol{\theta}^{\rm cyl})\\
        1+\overline{N}_{b|b}(\boldsymbol{\theta}^{\rm cyl})
        \end{pmatrix}\:.
\end{align}

\begin{figure*}[t]
    \centering
    \includegraphics[width=0.95\linewidth]{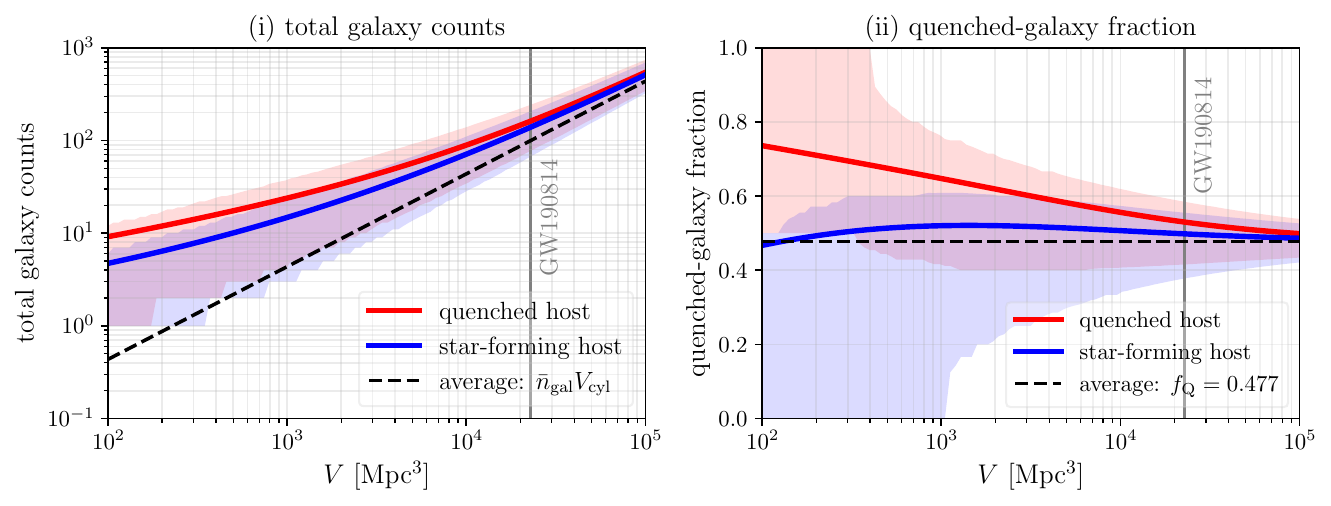}
    \caption{
    Host-conditioned galaxy environments for $A=5$.
    Curves show (i) the mean total count $\mu_{\rm Q}^{(h)}+\mu_{\rm SF}^{(h)}$ and (ii) the Q-galaxy fraction
    estimated from mean counts, $\mu_{\rm Q}^{(h)}/(\mu_{\rm Q}^{(h)}+\mu_{\rm SF}^{(h)})$, for Q (red) and SF (blue) hosts.
    Dashed lines indicate $\bar n_{\rm gal}V$ and $f_{\rm Q}$.
    Shading shows the central 68\% interval of a single host's counts and Q-galaxy fraction predicted by the PLN model.
    The vertical line marks the localization cylinder of GW190814, $V=2.3\times10^4\,{\rm Mpc}^3$.
    }
    \label{fig:sfq_environments}
\end{figure*}

\secl{Statistical distinguishability of host-galaxy properties} \label{sec:distinguishability} 
We describe the full counts-in-cylinders statistics by the host-conditioned distribution $p\!\left(\boldsymbol N(\boldsymbol\theta^{\rm cyl}) \mid\mathcal H_h,\boldsymbol\theta^{\rm cyl}\right)$, which can be calibrated from observed or mock galaxy catalogs. 
For our forecasts, we approximate the joint distribution of neighboring-galaxy counts by a bivariate Poisson--lognormal (PLN) model, motivated by its use in galaxy counts-in-cells analyses~\cite{2dFGRS:2004amf,DES:2016wzy}. 
Under $\mathcal H_h$, the host itself contributes a fixed count $\delta_{h,h'}$ to population $h'$ and therefore does not contribute to the count covariance.
The model approximately reproduces the central quantiles of the measured total-count distributions for the long cylinders tested.
Its construction, empirical comparison, and limitations are presented in the Supplemental Material.

For an individual event $i$, the relative support for an $a$-type over a $b$-type host is quantified by the likelihood ratio $\Lambda_i :=p(\boldsymbol N_i(\boldsymbol\theta_i^{\rm cyl}) \mid\mathcal H_a,\boldsymbol\theta_i^{\rm cyl}) / p( \boldsymbol N_i(\boldsymbol\theta_i^{\rm cyl}) \mid\mathcal H_b,\boldsymbol\theta_i^{\rm cyl})$.
Values $\Lambda_i>1$ favor an $a$-type host, whereas $\Lambda_i<1$ favor a $b$-type host. 
Values $\Lambda_i\simeq1$ provide little information about the host type.

To combine this event-level information into constraints on the host-population fractions of BBH mergers, we introduce $\lambda_a$, the fraction of selected events hosted by population $a$.
Weighting the two host hypotheses by their population fractions gives the single-event likelihood,
\begin{align}
\label{eq:population_count_mixture}
    p\!\left(
    \boldsymbol N_i(\boldsymbol\theta_i^{\rm cyl})
    \mid\lambda_a,\boldsymbol\theta_i^{\rm cyl}
    \right)
    &=
    \lambda_a\,p\!\left(
    \boldsymbol N_i(\boldsymbol\theta_i^{\rm cyl})
    \mid\mathcal H_a,\boldsymbol\theta_i^{\rm cyl}
    \right)
    \nonumber\\
    &\quad+
    (1-\lambda_a)\,p\!\left(
    \boldsymbol N_i(\boldsymbol\theta_i^{\rm cyl})
    \mid\mathcal H_b,\boldsymbol\theta_i^{\rm cyl}
    \right)\:.
\end{align}
For $N^{\rm ev}$ conditionally independent events, the population hyperlikelihood is~\footnote{
For this proof-of-principle application, we assume complete sampling of the adopted galaxy populations and that the hosts of the selected GW events belong to this sample.
These simplifying assumptions are not intrinsic to the framework, and GW-event and galaxy-catalog selection can be incorporated into the likelihood to infer intrinsic host-population fractions in future studies.
}
\begin{equation}
\label{eq:population_likelihood}
    \mathcal L_{\rm pop}(\lambda_a)
    =
    \prod_{i=1}^{N^{\rm ev}}
    p\!\left(
    \boldsymbol N_i(\boldsymbol\theta_i^{\rm cyl})
    \mid\lambda_a,\boldsymbol\theta_i^{\rm cyl}
    \right)\:.
\end{equation}
Given a hyperprior $\pi(\lambda_a)$, the posterior is
\begin{equation}
\label{eq:population_fraction_posterior}
    p\!\left(
    \lambda_a\mid
    \{\boldsymbol N_i(\boldsymbol\theta_i^{\rm cyl}),
    \boldsymbol\theta_i^{\rm cyl}\}
    \right)
    \propto
    \pi(\lambda_a)\,\mathcal L_{\rm pop}(\lambda_a)\:.
\end{equation}
Events favoring population $a$ support larger $\lambda_a$, and vice versa.
In the limiting case in which the host type of every event is identified exactly, with $N_a^{\rm ev}$ hosts of type $a$ and $N_b^{\rm ev}$ hosts of type $b$, Eq.~\eqref{eq:population_likelihood} reduces to the binomial form, $\mathcal L_{\rm pop}(\lambda_a)\propto \lambda_a^{N_a^{\rm ev}}(1-\lambda_a)^{N_b^{\rm ev}}$.
We use this framework below to quantify how increasingly informative GW localization environments combine to constrain the host-galaxy population.

\secl{Application to star-forming and quenched galaxies} 
\label{sec:sfq_application} 
We apply the framework to SF and Q galaxies in a volume-limited sample from Data Release 7 of the Sloan Digital Sky Survey (SDSS DR7)~\cite{2000AJ....120.1579Y,2009ApJS..182..543A}.
These common galaxy populations provide a simple test case for probing the connection between BBH mergers and host star-formation activity.
Following Ref.~\cite{Watson:2014zuk}, we select galaxies with $0.02<z<0.067$, $k+e$-corrected $M_r<-19$, and $\log_{10}(M_\star/M_\odot)>9.8$, separating SF and Q galaxies at a specific star-formation rate of $10^{-11}\,{\rm yr}^{-1}$. 
The number fractions are $f_{\rm SF}=0.523$ and $f_{\rm Q}=0.477$, with a mean galaxy number density $\bar n_{\rm gal}=4.36\times10^{-3}\,{\rm Mpc}^{-3}$.

We model the auto-correlations of the full, SF and Q samples with standard power-law fits to the measured projected clustering \cite{Davis:1982gc,2011ApJ...736...59Z} and obtain their SF--Q cross-correlation from the binary-partition identity, as detailed in the Supplemental Material. 
These models are extrapolated beyond the fitted separation range when evaluating the galaxy-count statistics. 
We fix the aspect ratio to $A=5$, approximately corresponding to the median aspect ratio for well-localized BBH events to date~\footnote{ 
The median is computed from events with $90\%$ localization volumes smaller than $10^6\,{\rm Mpc}^3$, using the GWTC-5.0, GWTC-4.1, and GWTC-2.1 public parameter estimation data and O4c public alerts in GraceDB \cite{LVK:GWTC5Data,LVK:GWTC4p1Data,LV:GWTC2p1PEData, Moe:2014GraceDB}. 
}. 
At this fixed ratio, we use the cylindrical volume $V$ in place of $\boldsymbol{\theta}^{\rm cyl}$.

Figure~\ref{fig:sfq_environments} shows the resulting host-conditioned galaxy environments. 
The shaded regions show the scatter in galaxy counts and Q-galaxy fractions among cylinders of the same volume and host type, arising from galaxy clustering and Poisson shot noise in the adopted PLN model. 
Consequently, the two host-conditioned distributions can overlap substantially, with some Q-host realizations exhibiting relatively SF-rich environments and vice versa. 
Q hosts nevertheless have both higher mean total counts and a larger Q-galaxy fraction than SF hosts. 
Even around SF hosts, the Q-galaxy fraction slightly exceeds $f_{\rm Q}$ at intermediate volumes because Q galaxies dominate the clustered excess. 
These host-dependent differences become increasingly apparent as the volume decreases below $10^4\,{\rm Mpc}^3$, while the environments approach the cosmic-mean galaxy population at larger volumes.
GW localizations are approaching these scales: the localization cylinder of GW190814 has $V=2.3\times10^4\,{\rm Mpc}^3$~\footnote{From its 90\% sky area $\Delta\Omega_{90}=18.5\,{\rm deg}^2$ and the 90\% credible interval on the luminosity distance, $D_{\rm L}=241^{+41}_{-45}\,{\rm Mpc}$~\cite{LIGOScientific:2020zkf}.
We use GW190814 only as an example of an achieved GW localization scale, irrespective of the uncertain nature of its secondary compact object.}.
Particularly well-localized events with distinctive galaxy environments may constrain individual host-galaxy types even before strong population-level constraints become available. 
Such opportunities should increase as detector sensitivities improve and more detectors operate jointly.

An important consequence of galaxy clustering is that the characteristic localization volume for individual host identification is substantially smaller than uniform-density estimates suggest.
The commonly adopted density $\bar n_{\rm gal}=0.01\ {\rm Mpc}^{-3}$ gives a characteristic volume of $100\,{\rm Mpc}^3$ (e.g.,~\cite{Chen:2016tys,Nishizawa:2016ood,Mo:2024frl}), but the left panel of Fig.~\ref{fig:sfq_environments} shows several candidate galaxies on average within this volume under either host hypothesis.
We define the characteristic host-resolvable volume by $\sum_{h'\in\{{\rm SF},{\rm Q}\}} \overline{N}_{h'|h}(V_{\rm res}^{(h)})=1$, corresponding to a mean of one neighboring galaxy around the host.
For $A=5$, this criterion gives $V_{\rm res}^{(\rm SF)}=7.4\,{\rm Mpc}^3$ and $V_{\rm res}^{(\rm Q)}=0.46\,{\rm Mpc}^3$, one and two orders of magnitude below this benchmark, respectively.
Both rely on extrapolating the fitted correlations to smaller separations and should be regarded as indicative, but the trend is general: positive clustering reduces the volume containing one expected neighbor relative to a uniform population of the same mean density.

\begin{table}[t]
    \centering
    \caption{
    Expected numbers of BBH events in one year satisfying $\sum_{h'\in\{{\rm SF},{\rm Q}\}} \overline{N}_{h'|h}(\boldsymbol{\theta}^{\rm cyl}_i)<1$ under each host hypothesis. 
    We evaluate the expected neighbor count for each event using its localization cylinder ${\boldsymbol{\theta}}_i^{\rm cyl}$.}
    \label{tab:forecast_resolvable_events}
    \begin{tabular*}{\columnwidth}{@{\extracolsep{\fill}}lcc@{}}
        \hline\hline
        Network & $h=\rm Q$ & $h=\rm SF$ \\
        \hline
        O5: HLVK        & $0.00$ & $0.07$ \\
        Post-O5: HLIVK  & $0.12$ & $0.21$ \\
        3G: ET+2CE      & $0.70$ & $3.24$ \\
        \hline\hline
    \end{tabular*}
\end{table}

\begin{figure}
    \centering    
    \includegraphics[width=0.95\linewidth]{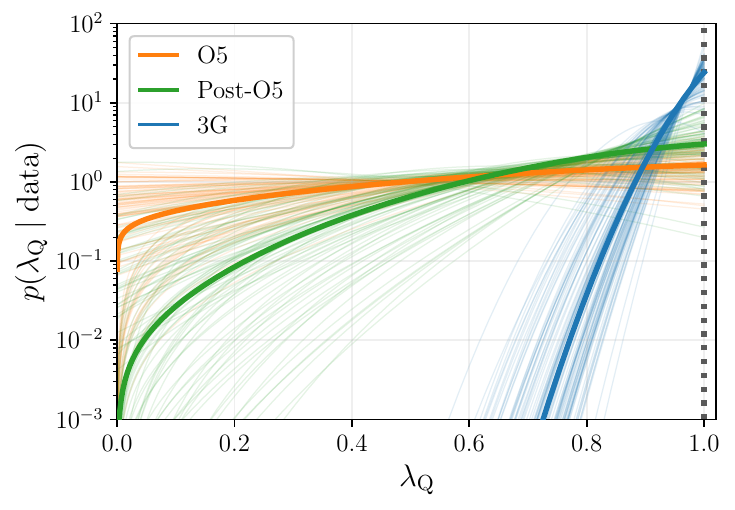}
    \caption{
    Forecast posterior distributions for the Q-host fraction $\lambda_{\rm Q}$ after one year of observation, assuming an all-Q host population with $\lambda_{\rm Q}=1$. 
    Orange, green, and blue curves correspond to the O5, post-O5, and 3G detector networks, respectively. 
    Thin curves show posteriors from individual mock realizations, while thick curves show posteriors constructed from the log likelihood averaged over mock realizations. 
    The vertical dotted line marks the injected value $\lambda_{\rm Q}=1$.
    }
    \label{fig:fq_posterior}
\end{figure}

\secl{Forecasts for distinguishing SF/Q host galaxies}
\label{sec:sfq_forecasts} 
We forecast one year of observations for three detector networks using the Fisher-matrix approximation.
We consider O5 with LIGO Hanford and Livingston, Virgo, and KAGRA, post-O5 with the same detectors plus LIGO-India, and a third-generation (3G) network consisting of the Einstein Telescope and two Cosmic Explorer detectors.
We assume continuous operation of all detectors.
The Fisher calculation, adopted BBH population, and merger-rate normalization are described in the Supplemental Material.
We retain simulated BBH events at $z\leq1$ with network $\mathrm{SNR}\geq12$ and cylindrical localization volumes $V\leq10^{8}\,{\rm Mpc}^3$, and restrict the sample to a sky fraction $f_{\rm sky}=0.34$, comparable to the coverage of the Dark Energy Spectroscopic Instrument (DESI)~\cite{Hahn:2022dnf}.
For each event $i$, the 90\% sky and distance uncertainties determine the cylindrical geometry $\boldsymbol{\theta}_i^{\rm cyl}$ used to evaluate its host-conditioned galaxy-count distribution.

Within this selection, we assume complete sampling of the adopted SF/Q populations and hold their local number densities and correlation functions fixed, neglecting redshift evolution and clustering-parameter uncertainties.
These are idealized catalog assumptions, as the forecast does not model the detailed DESI mask or magnitude-dependent selection. 
For each network, we generate 100 mock realizations, drawing Poisson-distributed event numbers and sampling localization geometries. 
Galaxy counts are drawn from the host-conditioned Poisson--lognormal distributions and analyzed with the same model. 

Table~\ref{tab:forecast_resolvable_events} shows the expected annual numbers of events with a mean neighbor count below unity, evaluated using each event's cylindrical localization geometry under each host hypothesis, so that no fiducial aspect ratio enters. 
For O5 and post-O5, these numbers are well below one per year, and even 3G network yields only $\mathcal{O}(1)$ events per year. 
Although this mean-count criterion does not guarantee unique host identification, it illustrates the scarcity of events reaching the relevant localization scales.

Even when individual hosts remain unresolved, their galaxy environments can constrain the BBH merger host-galaxy population. 
Figure~\ref{fig:fq_posterior} shows the one-year forecast posteriors for an all-Q host population with $\lambda_{\rm Q}=1$, adopting a uniform prior on $\lambda_{\rm Q}\in[0,1]$. 
The median one-sided 90\% lower credible limits on $\lambda_{\rm Q}$ across mock realizations are $0.189$, $0.476$, and $0.920$ for O5, post-O5, and 3G networks, respectively.
The posteriors become progressively more constraining from O5 to 3G network.

\secl{Conclusion} 
\label{sec:conclusion} 
We presented an observation-driven framework for probing the host-galaxy properties of BBH mergers through galaxy clustering, without requiring unique identification of individual hosts.
Clustering both increases the number of candidate galaxies and makes their abundance and composition informative about the host-galaxy type. 
Given the corresponding host-conditioned count distributions, the framework applies to common and rare galaxy populations defined by stellar mass, metallicity, morphology, AGN activity, or membership in galaxy clusters. 

For SF and Q galaxies, the characteristic volumes containing one expected neighbor are $7.4$ and $0.46\,{\rm Mpc}^{3}$, respectively, under our adopted clustering extrapolation and geometry. 
Despite these demanding scales, our proof-of-principle forecasts yield median one-sided 90\% lower credible limits on $\lambda_{\rm Q}$ of $0.19$, $0.48$, and $0.92$ after one year of O5, post-O5, and 3G observations, respectively, for an all-Q host population. 
These forecasts illustrate the potential for host-galaxy population inference before 3G era. 

By evaluating host-conditioned galaxy environments, the framework complements ensemble GW--galaxy cross-correlation studies (e.g.,~\cite{Raccanelli:2016cud,Scelfo:2018sny,Calore:2020bpd,Zazzera:2024agl,Chuman:2025ezp}) and catalog-based dark-siren inference~\cite{Gray:2026loe,Agapito:2026pbb}. 
It can constrain host-galaxy types for particularly well-localized events and host-population fractions across multiple events, without prescribing BBH formation efficiencies or merger delay-time distributions, thereby providing an independent probe of the astrophysics investigated through merger-rate-based approaches~\cite{Vijaykumar:2023bgs,Schiebelbein-Zwack:2026sqj}.

The forecasts assume host-centered localization volumes, complete sampling of the adopted galaxy populations, an extrapolated analytic clustering model without redshift evolution, and the same count model for generating and analyzing mock GW  data. 
Applications to observed events will require the full GW localization likelihood, survey selection, galaxy-property uncertainties, and validated or empirically calibrated host-conditioned count distributions. 
Joint analyses of source and host-galaxy properties, including individual mergers with distinctive masses or spins, could test the connections between mergers and galaxies predicted by competing formation scenarios.

\secl{Data availability} 
The galaxy sample is drawn from the publicly available SDSS DR7 \cite{2000AJ....120.1579Y,2009ApJS..182..543A}, with stellar masses and star-formation rates taken from the MPA-JHU value-added catalog \cite{SDSS:2002vxr,Brinchmann:2003db,2007ApJS..173..267S}.
The codes to reproduce the results can be found in~\cite{Watarai:2026GalaxyClusteringBBH}.

\secl{Acknowledgments}
We thank Masamune Oguri for useful discussions.
We acknowledge the use of Claude (Anthropic) and Codex (OpenAI) for assistance with numerical analysis, code development, and validation.
D.~W. is supported by the Japan Society for the Promotion of Science (JSPS) through the Overseas Research Fellowships program.
H.~I. is supported by JST SPRING, Grant No.~JPMJSP2108.

\section{Supplemental Material}
\setcounter{equation}{0}
\renewcommand{\theequation}{S\arabic{equation}}

\subsection{Real-space correlation-function fit}
\label{app:clustering_fit}
We require the real-space correlation functions of SF and Q galaxies, whereas the SDSS measurements provide projected correlation functions $w_{\mathrm{p}}(r_{\mathrm{p}})$. 
The parent sample is the NYU-VAGC \texttt{bright0} catalog~\cite{SDSS:2004dnq}, restricted to $0.02<z<0.067$ and $k+e$ corrected $M_r<-19$, matched within $1.5''$ to the MPA-JHU value-added catalogs~\cite{SDSS:2002vxr,Brinchmann:2003db,2007ApJS..173..267S} with $\log_{10}(M_\star/M_\odot)>9.8$, and split into SF and Q at $\log_{10}({\rm sSFR}/{\rm yr}^{-1})=-11$.
We first validate our clustering pipeline by reproducing the published measurements of Ref.~\cite{Hearin:2013foa} for the full parent sample and Ref.~\cite{Watson:2014zuk} for the SF and Q subsamples. 
The median ratios of our measurements to the published values over the 15 $r_{\mathrm{p}}$ bins are 1.0076, 1.0066, and 1.0008 for the full, SF, and Q samples, respectively, demonstrating sub-percent agreement in the median measurement ratios. 
The largest deviations, approximately 3--5\%, occur in the outermost bin.

\begin{figure}
    \centering
    \includegraphics[width=0.95\linewidth]{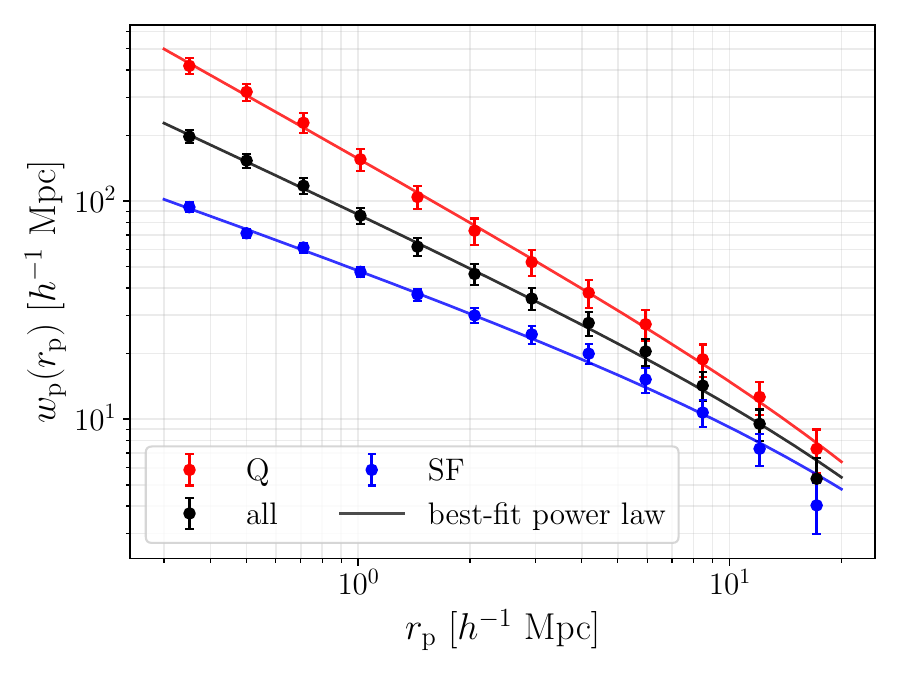}
    \caption{
    Projected two-point correlations $w_{\mathrm{p}}(r_{\mathrm{p}})$ for the full, SF, and Q galaxy samples. 
    Points show our measurements in the 12 bins used for fitting, with published uncertainties adopted as described in the text. 
    Solid lines show the finite-line-of-sight projections of the best-fitting real-space power-law models.
    }
    \label{fig:fit_w}
\end{figure}

We approximate each real-space auto-correlation function by a power law, following the standard empirical description of galaxy clustering (e.g.,~\cite{Davis:1982gc}),
\begin{equation}
    \xi_{a,a}(r)
    =
    \left(\frac{r}{r_{0,a}}\right)^{-\gamma_a},
    \qquad
    a\in\{{\rm all,SF,Q}\}\:.
\end{equation}
To account for the finite line-of-sight integration used in the measurement, we forward-project the  model as
\begin{equation}
    w_{\mathrm{p},a}^{\rm model}(r_{\mathrm{p}})
    =
    2\int_0^{40\,h^{-1}{\rm Mpc}} \diff\pi\,
    \xi_{a,a}\!\left(\sqrt{r_{\mathrm{p}}^2+\pi^2}\right)\:,
\end{equation}
with $h=0.7$.
This approximation neglects residual redshift-space distortions associated with the finite integration limit~\cite{vandenBosch:2012nq}, so $(r_{0,a},\gamma_a)$ are effective parameters of an approximate real-space model. 
We determine these parameters by fitting our measured projected correlations over $0.3<r_{\mathrm{p}}<20\,h^{-1}{\rm Mpc}$, minimizing a $\chi^2$ constructed from the diagonal errors reported in Table~1 of Ref.~\cite{Hearin:2013foa} for the full sample and Tables~A1 and A2 of Ref.~\cite{Watson:2014zuk} for the SF and Q subsamples. 
Correlations between separation bins are neglected in the fit. 
The resulting power laws are extrapolated outside this range where the volume-averaged quantities below require it. 
Table~\ref{tab:clustering_fit} summarizes the best-fit parameters, and Fig.~\ref{fig:fit_w} shows the measured projected correlation functions along with the projections of the best-fit power-law models. 
Over the fitted range, the quenched population exhibits systematically stronger clustering than the star-forming population, with the full sample lying between the two.

\begin{table}[h]
    \centering
    \caption{
    Best-fitting parameters of the power-law real-space auto-correlation functions, $\xi_{a,a}(r)=(r/r_{0,a})^{-\gamma_a}$. 
    The models are forward-projected to $\pi_{\max}=40\,h^{-1}{\rm Mpc}$ and fitted over $0.3<r_{\mathrm{p}}<20\,h^{-1}{\rm Mpc}$.
    }
    \label{tab:clustering_fit}
    \begin{ruledtabular}
    \begin{tabular}{lcc}
        Sample & $r_0\,[h^{-1}{\rm Mpc}]$ & $\gamma$ \\
        \hline
        All & 5.97 & 1.78 \\
        SF  & 4.61 & 1.59 \\
        Q   & 7.36 & 1.95 \\
    \end{tabular}
    \end{ruledtabular}
\end{table}

The SF and Q samples form a disjoint and exhaustive partition of the parent sample, so their auto- and cross-correlation functions satisfy the exact binary-partition identity~\cite{2011ApJ...736...59Z}. 
Inserting the independently fitted auto-correlations gives the model estimate
\begin{equation}
    \xi_{\rm SF,Q}(r)
    =
    \frac{
    \xi_{\rm all,all}(r)
    -f_{\rm SF}^2\xi_{\rm SF,SF}(r)
    -f_{\rm Q}^2\xi_{\rm Q,Q}(r)
    }{2f_{\rm SF}f_{\rm Q}}\:,
\end{equation}
where $f_{\rm SF}=0.523$ and $f_{\rm Q}=0.477$ are the number fractions in the parent sample. 
Although the identity is exact, the reconstructed cross-correlation inherits the approximations of the fitted auto-correlations and is not generally a power law. 
We use the fitted auto-correlations and this cross-correlation to compute the volume-averaged clustering entering the host-conditioned galaxy-count likelihood.

\subsection{Poisson--lognormal model for host-conditioned galaxy counts}
\label{app:pln}
For our forecasts, we model the joint distribution of neighboring-galaxy counts as bivariate Poisson--lognormal (PLN)~\cite{2dFGRS:2004amf,DES:2016wzy}, which allows for overdispersion and for correlations between the two populations.
Writing the volume-averaged two-point function as
\begin{equation}
    \Xi_{h',h''}(\boldsymbol{\theta}^{\rm cyl})
    :=
    \frac{1}{V^2}
    \int_V \diff^3\boldsymbol{x}_1
    \int_V \diff^3\boldsymbol{x}_2\,
    \xi_{h',h''}\!\left(|\boldsymbol{x}_1-\boldsymbol{x}_2|\right)\:,
    \label{eq:Xi_pln}
\end{equation}
we adopt the multiplicative covariance closure
\begin{equation}
    \Sigma_{h',h''}^{(h)}
    =
    \delta_{h',h''}\overline N_{h'|h}
    +
    \overline N_{h'|h}\,\overline N_{h''|h}\,\Xi_{h',h''}\:,
    \label{eq:pln_covariance}
\end{equation}
in which the central host is fixed under each hypothesis and contributes no
variance.
This closure is neither implied by the PLN assumption nor exact, since the host-conditioned covariance also involves host--galaxy--galaxy three-point correlations, which can be less important for counts projected over line-of-sight lengths much longer than the correlation length~\cite{Oguri:2014dqa}.
We adopt it because it approximately reproduces the measured variances for the long cylinders tested below.

The model is then
\begin{gather}
    N_{h'}-\delta_{h,h'}\mid\boldsymbol{x} \sim
    {\rm Poisson}\!\left[
    \overline N_{h'|h}\,e^{\,x_{h'}-S_{h',h'}/2}\right]\:,\\
    \boldsymbol{x}\sim\mathcal N_2(\boldsymbol 0,\boldsymbol S)\:,
    \label{eq:pln_model}
\end{gather}
with $S_{h',h''}=\ln(1+\Xi_{h',h''})$.
$\boldsymbol S$ is fixed by matching the PLN covariance to Eq.~\eqref{eq:pln_covariance}, and we verify that it is positive definite over the geometries to which the model is applied. 
The distribution then reproduces the specified mean and covariance by construction, while its higher moments, zero-count probabilities, and tails follow from the lognormal assumption rather than from the inputs.

\begin{figure*}[t]
    \centering
    \includegraphics[width=0.9\linewidth]{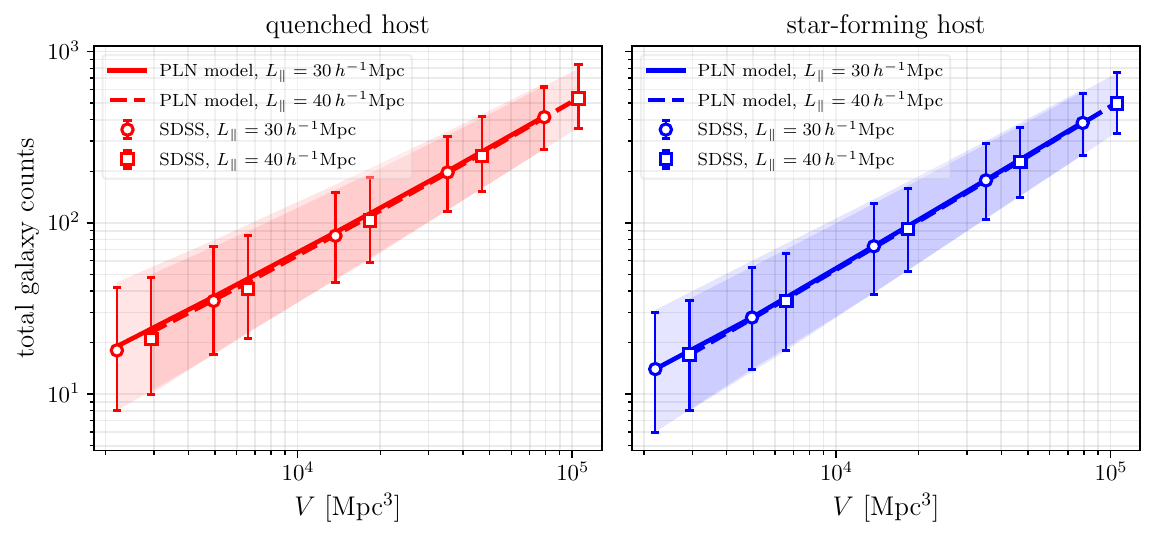}
    \caption{
    Total galaxy counts, including the central host, in SDSS host-centered cylinders for quenched (left) and star-forming (right) hosts. 
    Circles and squares show the measured medians for $L_\parallel=30$ and $40\,h^{-1}{\rm Mpc}$, respectively, with error bars spanning the 16th--84th percentiles. 
    Solid and dashed lines show the corresponding PLN medians, evaluated at the same cylinder geometries and connected separately for each $L_\parallel$. 
    Shaded bands show the model 16th--84th percentile ranges. 
    The error bars show the spread of the measured counts, not an uncertainty.
    }
    \label{fig:count_distribution}
\end{figure*}

We test the model against host-centered cylinder counts measured in SDSS over $2\leq R_\perp/(h^{-1}{\rm Mpc})\leq12$ and $2\leq L_\parallel/(h^{-1}{\rm Mpc})\leq40$, on the same sample as the correlation measurements. 
The counts are in redshift space, with comoving positions inferred from observed redshifts, and a cylinder is kept only if at least 99\% of 800 points drawn uniformly inside it fall within the survey mask and $0.02<z<0.067$, which retains 28--89\% of the central galaxies depending on the geometry, and 28--49\% for the cylinders shown in Fig.~\ref{fig:count_distribution}. 
Line-of-sight smearing by peculiar velocities suppresses the measured $\widehat{\Xi}$ relative to the real-space model, by a factor of 1.4--2.3 at $L_\parallel=2\,h^{-1}{\rm Mpc}$ but by at most 9\% for $L_\parallel\geq30\,h^{-1}{\rm Mpc}$, the lengths used in the comparison below.

We estimate $\Xi$ from the counts in cylinders centered on random points, as their covariance with the shot-noise term subtracted and normalized by the product of the mean counts. 
Combining this with the measured host-conditioned means isolates the closure from the fitted correlation functions, with no real-space model entering.
It reproduces the measured variances for $L_\parallel=30$ and $40\,h^{-1}{\rm Mpc}$ in 17 of the 20 host--geometry combinations within the $1\sigma$ jackknife uncertainties, the exceptions being quenched hosts at $40\,h^{-1}{\rm Mpc}$ where the measurement exceeds the prediction by up to 34\% ($1.9\sigma$); using the unconditional means $\bar n_{h'}V$ instead underestimates the same variances by factors of 1.3--8, whereas for $L_\parallel\leq20\,h^{-1}{\rm Mpc}$ the closure overestimates them, the ratio of measured to predicted variance falling from 0.74--0.89 at $20\,h^{-1}{\rm Mpc}$ to 0.18--0.63 at $2\,h^{-1}{\rm Mpc}$.

Figure~\ref{fig:count_distribution} tests the full model, with means and covariances from the fitted correlation functions, at the ten geometries $R_\perp\in\{2,3,5,8,12\}$ and $L_\parallel\in\{30,40\}\,h^{-1}{\rm Mpc}$. 
Over $2\times10^3\lesssim V/{\rm Mpc}^3\lesssim10^5$ the 16th, 50th, and 84th percentiles agree with the measurements to within about 10\%, with 51 of the 60 comparisons inside the $1\sigma$ jackknife uncertainties, although this tests only the central quantiles of the total count and not the joint SF/Q distribution or the likelihood ratios built from it.

\subsection{Fisher forecasts for future GW observations}
We use \texttt{GWFish}~\cite{Dupletsa:2022scg} to forecast the
cylindrical localization geometry $\boldsymbol{\theta}^{\rm cyl}=(R_\perp,L_\parallel)$ of each simulated BBH merger.
For the O5 network, LIGO Hanford and Livingston use the LIGO O5c sensitivity~\cite{LIGO:T2500310}, Virgo uses the O5 Tier-1 high-sensitivity curve~\cite{Virgo:VIR0797A25}, and KAGRA uses a curve with a 30\,Mpc binary neutron star range, obtained by rescaling the 128\,Mpc curve~\cite{LIGO:T2000012}.
For the post-O5 network, the three LIGO detectors, including LIGO-India, use the A$^\sharp$ baseline sensitivity~\cite{LIGO:T2300041}, Virgo uses the same O5 Tier-1 high-sensitivity curve, and KAGRA uses the \texttt{BB40FIS\_HQS} configuration~\cite{KAGRA:2025dra}. 
For 3G network, we use the ET-D sensitivity~\cite{Hild:2010id} for a 10\,km triangular Einstein Telescope and the Cosmic Explorer sensitivities~\cite{Evans:2021gyd} for two L-shaped detectors with 40\,km and 20\,km arms at the fiducial CE-A and CE-B locations of Ref.~\cite{Gupta:2023lga}, respectively.
All detectors are assumed to operate continuously.

In the high-SNR limit, assuming stationary Gaussian noise, the likelihood can be approximated as a multivariate Gaussian in the source parameters around its maximum.
The inverse of the Fisher matrix then approximates the corresponding parameter-estimation covariance,
\begin{equation}
    \Sigma = \Gamma^{-1}\:,
    \qquad
    \Gamma_{ij}
    =
    \sum_{d}
    \left(\partial_i\tilde h_d \,\middle|\, \partial_j\tilde h_d\right)_d\:,
\end{equation}
where $\tilde h_d(f)$ is the response of detector $d$,
$\partial_i:=\partial/\partial\theta_i$, and
\begin{equation}
    \left(a|b\right)_d
    :=
    4\,{\rm Re}
    \int_{f_{{\rm min},d}}^{f_{{\rm max},d}} \diff f\,
    \frac{a^*(f)\,b(f)}{S_{n,d}(f)}\:,
\end{equation}
with $S_{n,d}(f)$ the one-sided noise power spectral density of
detector $d$.
We take $f_{{\rm min},d}=8$\,Hz for LIGO O5, KAGRA, and Cosmic Explorer,
10\,Hz for Virgo, 3\,Hz for LIGO A$^\sharp$, and 2\,Hz for the Einstein
Telescope, and $f_{{\rm max},d}=1024$\,Hz except for the Einstein
Telescope, for which it is 2048\,Hz.

We adopt the \texttt{IMRPhenomXHM} waveform model for quasi-circular, aligned-spin BBH coalescences, including higher-order multipole modes~\cite{Garcia-Quiros:2020qpx}.
The parameter set is
\begin{equation}
    \{\theta_i\}
    =
    \{
    m_{1,2}^{\rm src},\chi_{1,2},
    t_{\rm c},\phi_{\rm c},\iota,\psi,
    \alpha,\delta,D_{\rm L}
    \}\:,
\end{equation}
comprising the source-frame component masses, aligned dimensionless spins, coalescence time and phase, inclination, polarization angle, right ascension, declination, and luminosity distance.
The detector-frame masses are $m_{1,2}^{\rm det}=(1+z)m_{1,2}^{\rm src}$.

We map the marginalized angular and distance uncertainties to a cylinder aligned with the line of sight.
The 90\% sky area and the corresponding transverse comoving radius are
\begin{align}
    \Delta\Omega_{90}
    &=
    2\pi\ln 10\,|\cos\delta|\,
    \sqrt{\det\Sigma_{(\alpha,\delta)}}\:,
    \\
    R_\perp
    &=
    D_{\rm C}
    \sqrt{\frac{\Delta\Omega_{90}}{\pi}}\:,
\end{align}
where $\Sigma_{(\alpha,\delta)}$ is the marginalized covariance submatrix for $\alpha, \delta$, $\Delta\Omega_{90}$ is expressed in steradians, and $D_{\rm C}$ is the comoving distance.
The line-of-sight half-length is obtained from the half-width of the central 90\% luminosity-distance interval,
\begin{equation}
    L_\parallel
    =
    1.645\,\sigma_{D_{\rm L}}
    \left|
    \frac{\diff D_{\rm C}}{\diff D_{\rm L}}
    \right|,
    \qquad
    \sigma_{D_{\rm L}}
    =
    \sqrt{\Sigma_{D_{\rm L}D_{\rm L}}}\:.
\end{equation}
For the adopted flat cosmology, $D_{\rm L}=(1+z)D_{\rm C}$ implies
\begin{equation}
    \frac{\diff D_{\rm C}}{\diff D_{\rm L}}
    =
    \left[
    1+z+\frac{H(z)D_{\rm C}(z)}{c}
    \right]^{-1}\:,
\end{equation}
where $H(z)$ is the Hubble expansion rate and $c$ is the speed of light.
The distance factors and Jacobian are evaluated at the fiducial source redshift, giving the cylindrical volume $V=2\pi R_\perp^2L_\parallel$.

We generate mock BBH mergers from the GWTC-5 Default BBH population model, with hyperparameters fixed to representative
values from the population inference~\cite{2026arXiv260527226T}, as listed in Table~\ref{tab:fiducial_bbh_population}.
We draw primary masses, mass ratios, redshifts, spin magnitudes,
and spin tilts from this model.
For the aligned-spin waveform calculation, we retain only the spin components parallel to the orbital angular momentum, neglecting precession.
We normalize the event yield using
\begin{equation}
    R_{\rm BBH}(z=0.2)
    =
    30.4\,{\rm Gpc}^{-3}\,{\rm yr}^{-1}\:,
\end{equation}
the median value of the reported rate~\cite{2026arXiv260527226T}.
The population hyperparameters and rate normalization are held fixed in these forecasts.

\begin{table*}[t]
\centering
\caption{
Fiducial hyperparameters adopted for the simulated BBH population.
Component masses are defined in the source frame.
}
\label{tab:fiducial_bbh_population}
\begin{ruledtabular}
\begin{tabular}{llc}
Parameter & Symbol & Fiducial value \\
\hline

\multicolumn{3}{c}{\textit{Primary-mass distribution}} \\
Low-mass power-law slope
    & $\alpha_1$
    & $1.5$ \\
High-mass power-law slope
    & $\alpha_2$
    & $5.4$ \\
Break mass
    & $m_{\rm break}$
    & $37.5\,M_\odot$ \\
Minimum primary mass
    & $m_{1,\min}$
    & $3.0\,M_\odot$ \\
Maximum primary mass
    & $m_{1,\max}$
    & $200\,M_\odot$ \\
Low-mass taper width
    & $\delta m_1$
    & $4.0\,M_\odot$ \\
Continuum mixture fraction
    & $\lambda_{\rm cont}$
    & $0.40$ \\
First Gaussian mixture fraction
    & $\lambda_{10}$
    & $0.55$ \\
First Gaussian mean
    & $\mu_{10}$
    & $10.0\,M_\odot$ \\
First Gaussian standard deviation
    & $\sigma_{10}$
    & $0.9\,M_\odot$ \\
Second Gaussian mixture fraction
    & $\lambda_{35}$
    & $0.05$ \\
Second Gaussian mean
    & $\mu_{35}$
    & $34.0\,M_\odot$ \\
Second Gaussian standard deviation
    & $\sigma_{35}$
    & $5.0\,M_\odot$ \\

\hline
\multicolumn{3}{c}{\textit{Mass-ratio distribution}} \\
Power-law index
    & $\beta_q$
    & $1.0$ \\
Minimum secondary mass
    & $m_{2,\min}$
    & $3.0\,M_\odot$ \\
Low-mass taper width
    & $\delta m_2$
    & $4.0\,M_\odot$ \\

\hline
\multicolumn{3}{c}{\textit{Spin distribution}} \\
Spin-magnitude mean
    & $\mu_a$
    & $0.08$ \\
Spin-magnitude standard deviation
    & $\sigma_a$
    & $0.35$ \\
Gaussian tilt-component fraction
    & $\zeta$
    & $0.65$ \\
Mean of $\cos\theta_i$
    & $\mu_{\cos\theta}$
    & $0.35$ \\
Standard deviation of $\cos\theta_i$
    & $\sigma_{\cos\theta}$
    & $1.2$ \\

\hline
\multicolumn{3}{c}{\textit{Redshift distribution}} \\
Redshift-evolution index
    & $\kappa_z$
    & $2.5$ \\

\end{tabular}
\end{ruledtabular}
\end{table*}

\bibliographystyle{apsrev4-2}
\bibliography{reference}

\end{document}